\documentclass[10pt, conference]{IEEEtran}

\usepackage{cite}
\usepackage{graphicx}
\usepackage[caption=false]{subfig}
\usepackage{amssymb, amsmath}

\usepackage[utf8]{inputenc} 
\usepackage[T1]{fontenc}
\usepackage{microtype}
\usepackage{mdframed}
\usepackage{tikz,pgfplots}
\usetikzlibrary{patterns}

\usepackage{multirow}

\mdfsetup{innerrightmargin=.7cm,
innerleftmargin=.7cm,
linecolor=white
}

\pgfplotsset{compat=newest}

\definecolor{myred}{RGB}{248,118,109}
\definecolor{mygree}{RGB}{124,174,0}

\begin{document}

\title{On the Interplay between Non-Functional Requirements and Builds on Continuous Integration}

\author{\IEEEauthorblockN{Kl{\'e}risson V. R. Paix{\~a}o\IEEEauthorrefmark{1},
Cr{\'i}cia Z. Fel{\'i}cio\IEEEauthorrefmark{2},
Fernanda M. Delfim\IEEEauthorrefmark{1} and
Marcelo de A. Maia\IEEEauthorrefmark{1}}
\IEEEauthorblockA{\IEEEauthorrefmark{1}Universidade Federal de Uberl{\^a}ndia --
Uberl{\^a}ndia (MG), Brazil\\
\{klerisson, fernanda, marcelo.maia\}@ufu.br}
\IEEEauthorblockA{\IEEEauthorrefmark{2}Instituto Federal do Tri{\^a}ngulo Mineiro --
Uberl{\^a}ndia (MG), Brazil\\
cricia@iftm.edu.br}
}

\maketitle

\begin{abstract}
Continuous Integration (CI) implies that a whole developer team works together on the mainline of a software project.
CI systems automate the builds of a software.
Sometimes a developer checks in code, which breaks the build.
A broken build might not be a problem by itself, but it has the potential to disrupt co-workers, hence it affects the performance of the team.
In this study, we investigate the interplay between non-functional requirements (NFRs) and builds statuses from 1,283 software projects.
We found significant differences among NFRs related-builds statuses. Thus, tools can be proposed to improve CI with focus on new ways to prevent failures into CI, specially for efficiency and usability related builds.
Also, the time required to put a broken build back on track indicates a bimodal distribution along all NFRs, with higher peaks within a day and lower peaks in six weeks.  Our results suggest that  more planned schedule for maintainability  for Ruby, and  for  functionality and reliability for Java would  decrease delays  related to broken builds.

\end{abstract}


\IEEEpeerreviewmaketitle

\section{Introduction}

\begin{mdframed}
``In general the answer to how to stay efficient when a build is almost always broken is: \emph{\textbf{stop breaking the build}}.'' -- \emph{Anonymous}\footnotemark

\end{mdframed}
\footnotetext{https://perma.cc/NS8Z-3GX8}

This excerpt from an online Question and Answer community lays out competing concepts of Continuous Integration (CI) in the software industry.
CI means that a whole developer team works together on the mainline of a software project~\cite{fowler2006continuous}.
CI build-process automatically takes source code commits, compiles the code, and then progresses through a pipeline of testing.
Sometimes one developer checks in the source code repository something that breaks the build,
i.e. checks in code which does not compile or pass unit or code analysis tests.
If on one hand, it may disrupt colleagues' work, on the other, it prevents breakages going unnoticed.

The build of a system is one of the first steps of moving software from development to customers. 
A failure in the build may not only disrupt the co-workers, but also the business~\cite{Kerzazi2014}. 
Hence, as important as avoiding broken builds is the time taken to fix the build.
Longer times mean more wasted developer time.
Understanding for what reasons a set of source code changes broke the build is hard without developer's advice and becomes crucial to prevent problems~\cite{AdamsMcIntosh2016}.
Also, relying on the developer for such analysis it is feasible on small-scale cases.

A growing body of work in software engineering uses \emph{topic analysis} to make sense of textual data in software repositories~\cite{Chen2016}.
As we gain access to larger datasets, it becomes important to scale our ability to conduct such analyses~\cite{DeSouza:2013}.
In this direction, Hindle~et~al. established a link between topics computed from commit messages and non-functional requirements (NFRs)~\cite{Hindle:2013}.
Their technique enables large-scale topic analysis over such artifacts, because NFRs are widely spread across software projects.
Furthermore, that work shed some light on what a set of commit messages means in terms of NFRs.
Therefore, \emph{if failed CI builds are related to certain NFRs, then developers can use topics to prevent failures}.

In this study, we examine the NFRs categories computed from the list of all commits that were built in a given build job from Travis-CI (a CI platform for open-source software development~\cite{msr17challenge}) in 1,283 projects from GitHub repository. By studying a large corpus of projects, we aim to empirically investigate the interplay of NFRs and Travis-CI builds statuses.

Our research is guided by two main research questions:
\begin{description}
\item[\textbf{RQ1.}] Which NFRs occur more frequently in failed Travis-CI builds than successful ones?
\item[\textbf{RQ2.}] How long do NFR-related builds remain broken?
\end{description}

We found significant differences among NFRs related-builds statuses. Thus, tools can be proposed to improve CI with focus on new ways to prevent failures into CI. Further, our results suggest that more planned schedule for maintainability for Ruby, and for functionality and reliability for Java would decrease delays related to broken builds.

The paper outline is standard: literature review, material and method description, results, and conclusions.

\begin{figure*}[!ht]
  \centering 
  \includegraphics[width=\textwidth]{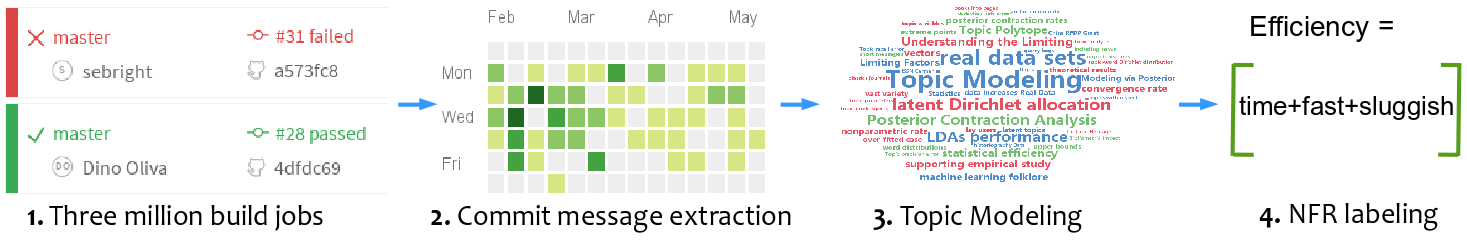} 
  \caption{The automatic non-functional requirement labeling computes commit messages topics from over 3 million Travis-CI build jobs.}
  \label{fig:method}
  \vspace{-1.5em}
\end{figure*}

\section{Related Work}

Our study inherits from a rich ecosystem of tools and applications for software repository mining, and draws on the insights of prior work in NFRs and topic modeling.

\emph{\textbf{Non-functional requirements.}}
While a functional requirement describes what a system should do, NFRs place constraints on the performance of the system, i.e. how it will do so~\cite{nfrs}.
NFRs may also describe aspects of the system related to its evolution over time, e.g., maintainability, extensibility, and documentation, to name a few. 
Unfortunately, there are a lot of disagreements on what NFRs really are.
Mairiza~et~al. found 114 different NFRs classes~\cite{Mairiza:2010:INN}, which contrasts with the international standard ISO 9126 quality model~\cite{iso}, where NFRs are defined by six high-levels classes: maintainability, functionality, portability, efficiency, usability, and reliability.
Eckhardt~et~al. analyzed  NFRs taken from industrial requirements specifications to better understand their nature~\cite{Eckhardt:2016:NRR}.
Their results suggest that NFRs are buried in functional requirements, insofar as we should not 
make any distinguish between them.
Despite the aforementioned discussion, whether NFRs are correctly categorized, we cannot deny that NFRs concepts pervade all modern software projects, therefore, we can use such definitions to compare projects.

\emph{\textbf{Topic Modeling.}}
The history of topic models in academic research related to Software Engineering is long.
For a comprehensive survey on this matter, we refer to Chen~et~al.~\cite{Chen2016}.
Herein, we focus on topic analysis applied to commit messages.
Hindle~et~al. mined commits in a windowed time fashion~\cite{Hindle2009}.
They applied \emph{latent dirichlet allocation} (LDA)~\cite{lda} technique in a 30-day period of commit messages to identify topic trends.
Their technique allows the automated summarization of ``what has been done'' in a given time.
In another work~\cite{hindle2010}, those authors also used topic analysis to annotate commit messages, among other software artifacts, and map the results onto software project phases.
Their idea is to propose an alternative approach to monitor software process compliance.
With respect to the study of CI builds statuses, there are some common themes between our work and Santos and Hindle's work~\cite{Santos:2016:JCC}.
In that work, a n-gram language model was proposed to compute how ``unusual'' is a commit message. 
The results suggest a positive correlation between unusualness messages and builds failures.

In comparison, our goal is to investigate the correlation between NFRs  developers were working on and CI builds statuses.
We rely on the method of Hindle~et~al. that links a set of commits messages to NFRs~\cite{Hindle:2013}.
However, since we are interested in system-wide builds statuses, our NFR related topics are extracted from all commits reported in CI builds.

\section{Material and Method}

\subsection{TravisTorrent Dataset}

TravisTorrent~\cite{msr17challenge} is a synthesis of software projects from GitHub that have Travis-CI enabled. Version 8.2.2017 comprehends 3,702,595 builds from 1,283 projects. For our particular interest, the structure of the build entries involves the job id, project name, status, builds duration, started timestamp, and all commits that were built.

Regarding the status of a build, there are five values in the dataset.
We consider in this study three of them:
\emph{passed}, which means a project has been built and passed its test suite;
\emph{failed}, a project failed to build or failed in its tests;
and \emph{errored}, a misconfiguration was found in the project.                   
The last two statuses were grouped. Ultimately, they both mean that the build is \emph{broken}.
We discard the other two statuses (started and canceled), because we either do not know the process outcome and the reasons behind its cancellation.

Additionally, with the project name, we fetch (clone) the repository from GitHub.
Then, with the commit list, all messages are taken.

\begin{table*}[!htbp]
\renewcommand{\arraystretch}{1.3}
\centering
 \caption{Pairwise Chi-Square comparison of NFRs. }
 \label{tab:chitest}
 \subfloat[Ruby projects.\label{tab:chi:ruby}]{
 \begin{tabular}{rccccc}
   				& \bfseries Portability &  &  &  &   \\ \cline{2-2} 
 \bfseries Usability 	   	& \multicolumn{1}{c|}{3.399654e-18}   &  \bfseries Usability &     &   &  \\ \cline{3-3}
 \bfseries Efficiency		& 2.919159e-09  & \multicolumn{1}{c|}{2.834015e-06}    &  \bfseries Efficiency   &    & \\ \cline{4-4}
 \bfseries Reliability	& 5.381500e-179 & 5.705839e-281 & \multicolumn{1}{c|}{1.193036e-260}    &  \bfseries Reliability    &     \\ \cline{5-5}
 \bfseries Maintainability& 4.084665e-29 & 1.919558e-40 & 8.737777e-37 & \multicolumn{1}{c|}{0.38027170}  &  \bfseries Maintainability \\ \cline{6-6}
 \bfseries Functionality & 0.000000e+00 & 0.000000e+00 & 0.000000e+00 & 0.01699222 & 0.05420935 \\ \cline{2-6}
 \end{tabular}
 }
 
 \vspace{-.5em}
 
 \subfloat[Java projects.\label{tab:chi:java}]{
 \begin{tabular}{rccccc}
  				& \bfseries Portability &  &  &  &   \\ \cline{2-2}
 \bfseries Usability 	   	& \multicolumn{1}{c|}{4.189831e-84}  & \bfseries Usability &  &   &   \\ \cline{3-3}
 \bfseries Efficiency		& 1.615018e-110 &  \multicolumn{1}{c|}{6.077951e-03}  &  \bfseries Efficiency &  &   \\ \cline{4-4}
 \bfseries Reliability	& 3.806572e-191 & 7.352944e-77  & \multicolumn{1}{c|}{5.284936e-72}  & \bfseries Reliability &  \\ \cline{5-5}
 \bfseries Maintainability& 4.564365e-87  & 1.090704e-39  & 6.556727e-37 & \multicolumn{1}{c|}{7.532970e-06} &  \bfseries Maintainability \\ \cline{6-6}
 \bfseries Functionality  &  5.678315e-12  & 9.714129e-188 & 9.305286e-240 & 7.173664e-288 &  1.335613e-117	\\ \cline{2-6}
 \end{tabular}
 }
 \vspace{-1.5em}
\end{table*}

\begin{figure*}[!htbp]
\centering
\subfloat[Ruby.\label{fig:ruby}]{
\begin{tikzpicture}[scale=1]
\begin{axis}[
	small,
    ybar,
    bar width=11pt,
    x=30pt,
    enlarge y limits={abs value=3,lower, upper},
    legend style={at={(.86,.97)}, anchor=north,font=\small},
    ylabel={\% builds},
    ymin=0,
    symbolic x coords={Maintainability,Functionality,Portability,Efficiency,Usability,Reliability,Unnamed},
    xtick=data,
    nodes near coords,
    nodes near coords align={vertical},
    x tick label style={rotate=35,anchor=east},
    every node near coord/.append style={font=\tiny},
    ]
\addplot[draw=black!50!green, pattern color = green, pattern = dots] coordinates {(Maintainability,0.27) (Functionality,9.38) (Portability,6.00) (Efficiency,27.27) (Usability,19.82) (Reliability,1.87) (Unnamed,2.5)};
\addplot[draw=red, fill=myred] coordinates {(Maintainability,0.10) (Functionality,3.51) (Portability,2.94) (Efficiency,13.7) (Usability,10.09) (Reliability,0.71)(Unnamed,1.76)};
\legend{Passed,Broken}
\end{axis}
\end{tikzpicture}
}
\subfloat[Java.\label{fig:java}]{
\begin{tikzpicture}[scale=1]
\begin{axis}[
	small,
    ybar,
    bar width=11pt,
    x=30pt,
    enlarge y limits={abs value=3,lower, upper},
    legend style={at={(.86,.97)}, anchor=north,font=\small},
    ylabel={\% builds},
    ymin=0,
    symbolic x coords={Maintainability,Functionality,Portability,Efficiency,Usability,Reliability,Unnamed},
    xtick=data,
    nodes near coords,
    nodes near coords align={vertical},
    x tick label style={rotate=35,anchor=east},
    every node near coord/.append style={font=\tiny},
    ]
\addplot[draw=black!50!green, pattern color = green, pattern = dots] coordinates {(Maintainability,0.36) (Functionality,7.75) (Portability,6.21) (Efficiency,34.85) (Usability,17.52) (Reliability,2.37) (Unnamed,4.14)};
\addplot[draw=red, fill=myred] coordinates {(Maintainability,0.23) (Functionality,1.94) (Portability,1.75) (Efficiency,13.34) (Usability,6.55) (Reliability,1.23)(Unnamed,1.69)};
\legend{Passed,Broken}
\end{axis}
\end{tikzpicture}
}
\caption{Passed vs. Broken builds. Figures on bars indicate percentages.}
\label{fig:contigency}
\vspace{-1.5em}
\end{figure*}
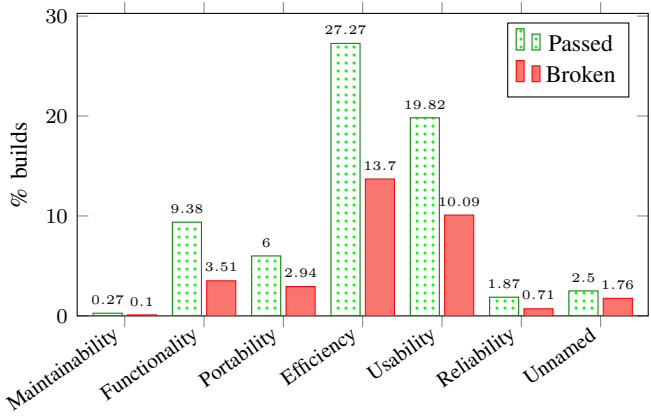
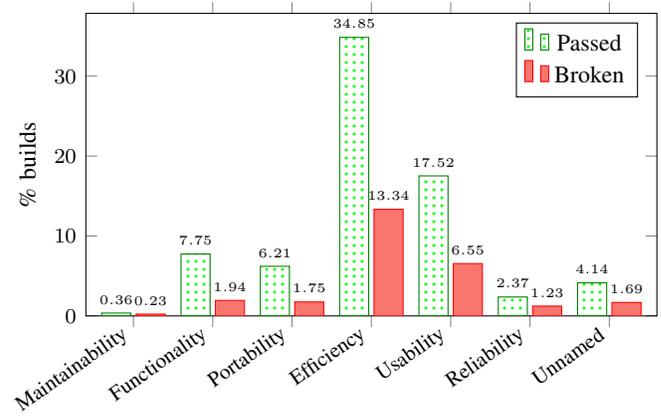

\subsection{NFR Labeling}\label{method}

The overall NFR automatic labeling process is illustrated as Fig.~\ref{fig:method}.
First, for each GitHub project we clone the repository.
Then, we select the commits that were built for each build job.
With the commit we fetch the associated messages.

Such messages, per project, are given as input to the topic modeling phase.
We use the Mallet toolkit~\cite{mallet}  to generate 20 topics with 10 words per topic.
To automatically label each build job with a topic, 
we use the \emph{exp3} word-list, please refer to the work of Hindle~et~al.~\cite{Hindle:2013} for the details on the word-list generation.
This word set consists of keywords separated by each NFR (maintainability, functionality, portability, efficiency, usability, and reliability).

The motivation to choose this word-list instead of others (exp1 or exp2) is because it contains more words per NRF category.
Since we aim to contrast diverse projects a broad list of words might be better representative.
Recall that this process is done per project. So, the topic computing of one project is not affected by the topics from others.

Finally, with each build job and its associated topic, we labeled our build job with an NFR where there was a match between the topic's word and the word-list.

\section{Results and Discussion}
This section reports our results.
For replication purposes, raw data used for our analyses is available for download\footnote{https://doi.org/10.6084/m9.figshare.2279505.v1}.

\subsection*{RQ1. Which NFRs occur more frequently in failed Travis-CI builds than successful ones?}

While a common best practice on continuous integration is to have all the tests passing at all times,
build breakage happens.
The primary endpoint of the study is to identify patterns of failure that might help developers prioritize their efforts on preventing such failures.

Fig.~\ref{fig:contigency} show, for each NFR and for different programming language (Ruby and Java), the percentage of passed and broken builds. Unnamed indicates the number of builds the approach was not able to classify automatically, as explained in Section~\ref{method}. Therefore, we do not consider this category in our statistical analysis.

To test the presence of a significant difference among proportions of builds we perform a Pearson Chi-Square pair-wise test on a contingency table, where columns represent the builds per NFR and rows builds statuses ($H0$: the proportion of builds having different statuses does not change among NFRs). P~values~<~0.05 were considered statistically significant. Table~\ref{tab:chitest} shows the P~values of paired NFRs.

For Ruby projects, analyses revealed significant differences in 10 out of the 15 pairwise comparisons.
There are no significant differences between \emph{efficiency} and \emph{portability} or \emph{usability}.
The same is observed with \emph{functionality} and \emph{reliability} or \emph{maintainability}.
With Java projects, all pair-wise comparisons were significant except between \emph{efficiency} and \emph{usability}, \emph{functionality} and \emph{portability}, and \emph{maintainability} and \emph{reliability}.

\subsection*{RQ2. How long do NFR-related builds remain broken?}

Here, we investigate the impact of broken builds considering the time elapsed until the build is fixed.
Although is not desirably to face build failures, they play an important role to the development process. For example, a broken build denotes a bug caught earlier~\cite{Hilton:2016}.
However, since the developers base their work on project branches, if they remain broken for longer times they affect the project's performance.

\begin{figure*}[!ht]
  \centering
  \subfloat[Ruby.\label{fig:boxruby}]{
  	\includegraphics[width=7.0cm, height=5.5cm]{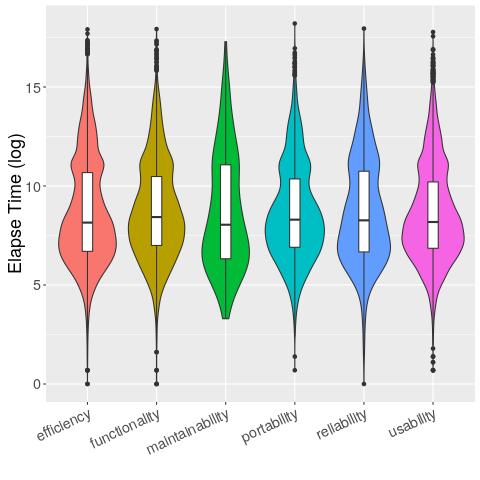} 
  }
  \hspace{1cm}
  \subfloat[Java.\label{fig:boxjava}]{
  	\includegraphics[width=7.0cm, height=5.5cm]{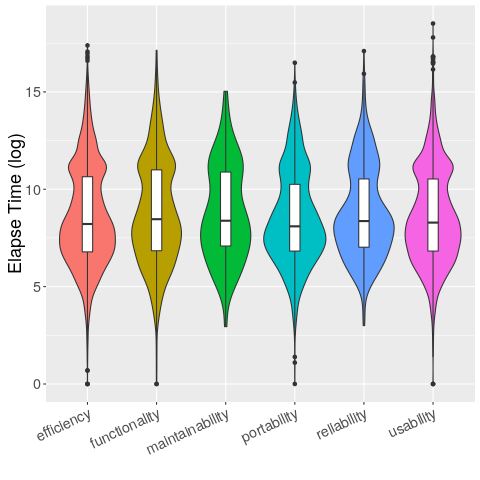} 
  }
  \caption{Graphical distribution of broken builds along the time group by NFR.}
  \label{fig:box}
  \vspace{-1em}
\end{figure*}

Table~\ref{tab:timeelapsed} shows the average time elapsed between a broken build and a sequent passed one group by NFR.
Fig.~\ref{fig:box} shows the graphical distribution of broken builds for each setting.

\begin{table}[!ht]
\renewcommand{\arraystretch}{1.3}
\centering
\caption{Average duration of broken builds in minutes.}
\label{tab:timeelapsed}
	 \begin{tabular}{rcc}
          \hline
        \multicolumn{1}{c}{\bfseries NFR} & \bfseries Ruby & \bfseries Java \\ 
        \hline
        Maintainability & 403 & 37 \\ 
        Functionality   & 144 & 58 \\ 
        Portability     & 121 & 34 \\ 
        Efficiency      & 118 & 40 \\ 
        Usability       & 73  & 36 \\ 
        Reliability     & 191 & 64 \\ 
        Unnamed         & 143 & 19 \\ 
        \hline 
        \bfseries Total average & 170 & 41 \\ \hline
     \end{tabular}
     \vspace{-1.5em}
\end{table}

\emph{\textbf{Discussion.}}
The goal of RQ1 was to examine whether providing comparison between NFR related builds statuses had an impact on continuous integration builds.
The study revealed significant results.
For Ruby projects, despite the absolute number of builds related to \emph{efficiency}, it holds the same proportion of passed and broken builds as \emph{usability} group.
Together they represent around 70\% of the builds. However, RQ2 results exposes that a broken build related to \emph{efficiency} NFR takes 1.6x more time on average to be fixed than a \emph{usability} broken build. 
We observe similar scenario for Java.
Broken builds from \emph{reliability} group has no significant proportional differences with \emph{maintainability} group, but an issue from the former group takes 1.7x more time on average regarding the last group.

Fig.~\ref{fig:box} shows the distribution of the time taken of broken build until a sequent successful build per NFR. Note that the distribution is bimodal along all NFRs. That is, it has two peaks, showing that most broken builds are either fixed within a day (higher peak) or takes around six weeks (lower peak), which is a common release methodology adopted in industry (rapid release cycles).  

Our design decisions suggest a set of limitations, many of which we hope to address in future work. We did not measure accuracy of the NFR labeling method.
Further, we study the association of builds with only one topic, but there might be cases where they can be linked with multiple topics.
Although our approach can be seen as a replication of the work Hindle~et~al.~\cite{Hindle:2013}, further evaluation is needed.
Finally, we only consider NFRs in our study. Thus, we refrain to only discuss about the relationship of NFRs and builds statuses. Future work could use/propose other classification of builds.

\section{Conclusion}
We examined a large set of projects to expose the relationship between NFR and CI builds statuses.
Certain categories of NFR related builds are more prevalent, such as efficiency and usability, regardless if Ruby or Java.
So, recommendation systems to help avoiding breakages on those kind of builds would produce overall larger impact on the whole process.  

Moreover, maintainability for Ruby projects, and functionality together with reliability for Java, take longer times to be fixed. So, they could be postponed to whenever developers are available to watch the builds, avoiding conflicts among themselves.

\noindent \emph{\textbf{Acknowledgment.}} We thank the Brazilian agencies {\small CAPES}, {\small CNPq}, and {\small FAPEMIG}.
\vspace*{-1.2em}

\bibliographystyle{IEEEtran}
\bibliography{biblio}

\end{document}